\documentclass[sigconf]{acmart}

\usepackage{booktabs} 
\usepackage[normalem]{ulem} 

\copyrightyear{2018} 
\acmYear{2018} 
\setcopyright{rightsretained} 
\acmConference[MM '18]{2018 ACM Multimedia Conference}{October 22--26, 2018}{Seoul, Republic of Korea}
\acmBooktitle{2018 ACM Multimedia Conference (MM '18), October 22--26, 2018, Seoul, Republic of Korea}
\acmDOI{10.1145/3240508.3241393}
\acmISBN{978-1-4503-5665-7/18/10}

\begin{document}
\title{SoniControl - A Mobile Ultrasonic Firewall}

\author{Matthias Zeppelzauer}

\affiliation{%
  \institution{St. P\"olten University of Applied Sciences}
}
\email{m.zeppelzauer@fhstp.ac.at}

\author{Alexis Ringot}
\affiliation{%
  \institution{St. P\"olten University of Applied Sciences}
}
\email{alexis.ringot@fhstp.ac.at}

\author{Florian Taurer}
\affiliation{
\institution{St. P\"olten University of Applied Sciences}
}
\email{florian.taurer@fhstp.ac.at}




\renewcommand{\shortauthors}{M. Zeppelzauer et al.}

\begin{abstract}

The exchange of data between mobile devices in the near-ultrasonic frequency band is a new promising technology for near field communication (NFC) but also raises a number of privacy concerns. We present the first ultrasonic firewall that reliably detects ultrasonic communication and  provides the user with effective means to prevent hidden data exchange. This demonstration showcases a new media-based communication technology (``data over audio") together with its related privacy concerns. It enables users to (i) interactively test out and experience ultrasonic information exchange and (ii) shows how to protect oneself against unwanted tracking. %
\end{abstract}

%
%



\maketitle

\section{Introduction}

Mobile multimedia data (especially audio) can be used aside from entertainment purposes also for near field communication (NFC). Several recently introduced technologies use (almost) inaudible near-ultrasonic signals (18-22kHz) to exchange information across the auditory channel through the loudspeakers and microphones of mobile devices  \cite{nearby,lisnr,shopkick}. This technology, also called \emph{data over audio}, enables effective near field communication capabilities but also opens up a number of security and privacy concerns. Data over audio can be used as a side channel to track users and their behavior across different devices without their knowledge~\cite{silverpush}. In 2015 already, the company Silverpush claimed to be tracking 18M devices with their ultrasonic technology \cite{silverpush18m}. A recent study has investigated the potential risk of ultrasonic side channels by systematically investigating more than 1.3 million mobile applications for ultrasonic tracking capabilities \cite{arp2017privacy}. The study identified 234 applications with ultrasonic tracking functionality where the most popular had between 1 and 5 million downloads.  Furthermore, research has shown that ultrasonic data exchange can be used as covert channel to exfiltrate and leak information from devices \cite{Do2015,Carrara2015,Sun2016}. Recently, the threat has been recognized by researchers and first ideas for defensive mechanisms have been proposed \cite{mavroudis2017privacy}. 

To prevent acoustic tracking and infiltration and to raise awareness to this novel technology, we have developed a mobile ultrasonic firewall that monitors the ultrasonic frequency band, acoustically detects different types of ultrasonic activity and notifies the user about the hidden data transfer. To effectively protect from unwanted ultrasonic data exchange, our firewall provides a jamming function that prohibits further communication. The SoniControl firewall\footnote{project webpage: \url{http://sonicontrol.fhstp.ac.at}}  presents a first means towards the effective protection against acoustic user- and cross-device tracking. In the following, we present the system design and a robust real-time algorithm for the detection and prevention of ultrasonic activity.

\section{System design}

\begin{figure}[t]
\includegraphics[trim={0 0cm 0 0cm},clip,width=1\linewidth]{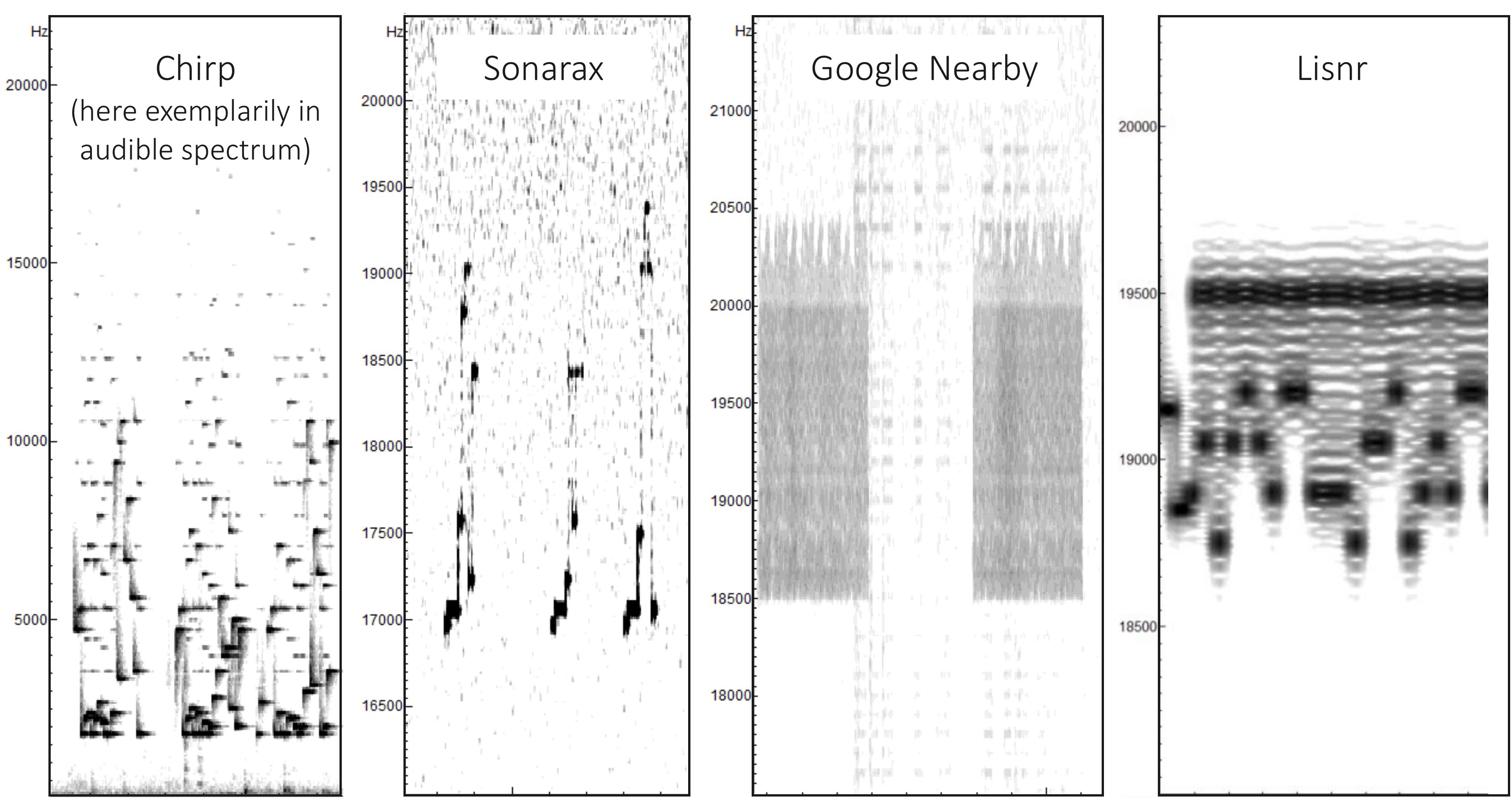} 
\caption{Different ultrasonic transmission technologies.}
\label{fig:technologies}
\end{figure}

\begin{figure}[ht]
\includegraphics[trim={0 0cm 0 0cm},clip,width=0.95\linewidth]{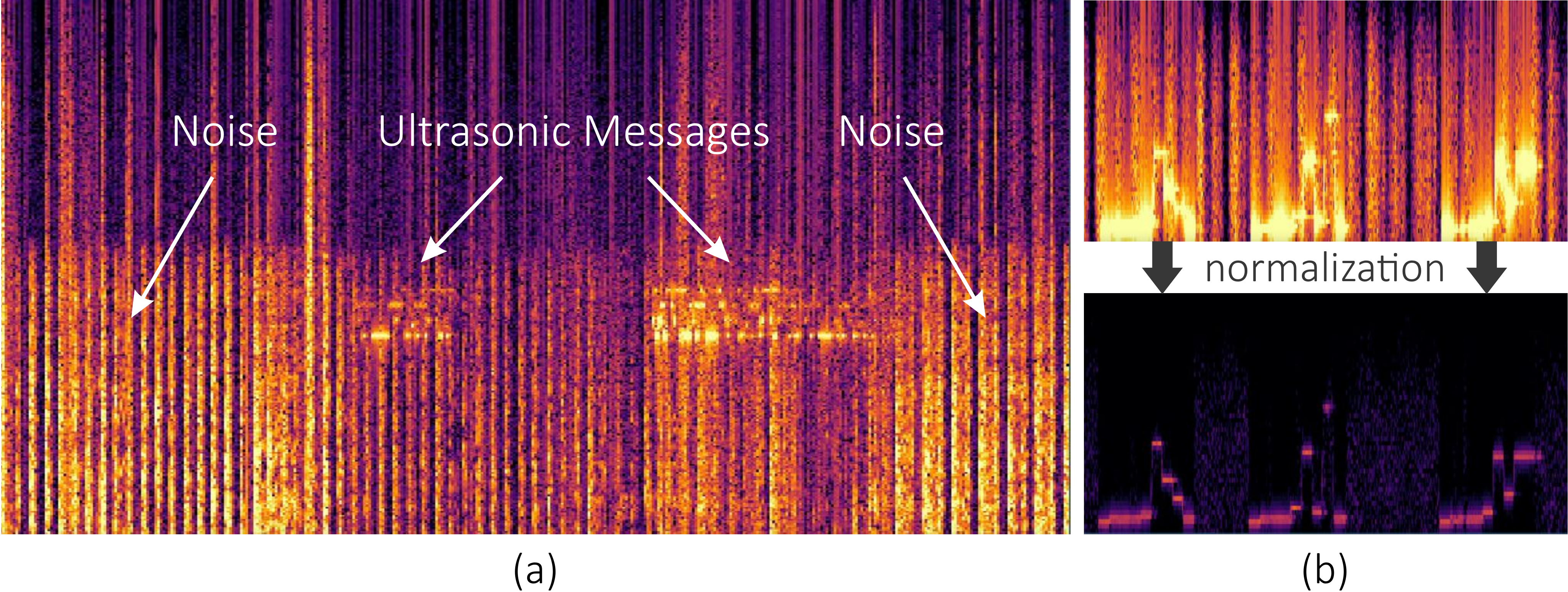} 
\caption{Ultrasonic noise: (a) ultrasonic messages in presence of noise; (b) noise removal by spectral normalization.}
\label{fig:noise}
\end{figure}

Different modulation and signal processing techniques such as frequency shift keying (FSK) and phase shift keying (PSK) can be used to encode information into an audio signal, see Figure \ref{fig:technologies} for  examples. The major challenge is the development of a detector that is technology-agnostic and still robust to noise. Noise goes beyond simple additive Gaussian noise, since sounds from the audible spectrum extend into the ultrasonic band, see also Figure \ref{fig:noise}(a). This renders pure energy-based detectors insufficient \cite{urkowitz1967energy, carrara2015characterizing}.

We model the spectral distribution in the ultrasound band statistically, build (and continuously update) a robust background model of the local surrounding and use this model to detect ultrasonic activity. In more detail, the signal detection is performed as follows. We first remove audible frequencies with a high-pass filter and compute the short-time FFT spectrum. We normalize the spectrum to remove broad-band noise, which frequently occurs in the near-ultrasound band. After normalization, most ambient noise is removed and the transmission frequencies become more salient, see Figure \ref{fig:noise}(b). Each input frame is stored in a cyclic buffer of a few seconds duration (e.g. 10s). Once the buffer is full, we compute the median of the spectral frequency distribution over time to obtain a robust estimate of the ambient sound distribution. To test for detections, we compute the Kullback-Leibler (KL) divergence between the background model and the spectral distribution of the incoming signal. The KL divergence indicates if both samples stem from the same distribution (value range 0 to 1). We apply a simple thresholding (threshold $t$) to the KL divergence to detect changes in the ultrasonic spectrum. Additionally, we use a running median filter to track if changes in the spectrum sustain over longer time. This avoids false detections and makes the system more robust. The detector has only one parameter to tune (threshold $t$) which has shown to be rather insensitive (0.5 yielded good results in all experiments so far).

When a detection is found, an alert is shown to the user. She can decide to let the communication pass or to actively block it to avoid further data transfer. For each detection, we store the time and location locally on the user's device for future reference. Places where the user decided to block the communication can automatically be blocked on the next visit. To block signals, we have implemented a flexible noise generator, which produces white noise at selected frequencies to jam different ultrasound protocols. For broadband signals like Google Nearby (see Figure \ref{fig:technologies}) a broadband signal is generated. To avoid any type of ultrasonic communication, the user can switch on \emph{preventive blocking}, which starts jamming immediately when a signal is detected.

To reduce the power consumption and  unnecessary noise pollution, as an alternative to acoustic blocking, we first try to obtain exclusive access to the microphone to avoid other apps listening to ultrasound. Note that this does not affect voice telephony.

\subsection{Implementation}

The system has been implemented for Android 4.1 and above and uses the Superpowered library for audio processing \cite{superpowered}. It requires a minimum of permissions (microphone and optionally location) and has been highly optimized to reduce computing power (CPU consumption $\le15$\% on a Samsung Galaxy S5 phone. Source code is available at: \url{https://git.nwt.fhstp.ac.at/m.zeppelzauer/SoniControl}.

\subsection{Experiments}
Experiments with a dataset of ultrasonic sound recordings\footnote{available at: \url{http://sonicontrol.fhstp.ac.at}} have shown that the detection algorithm is able to robustly detect different transmission protocols, including Google Nearby, Lisnr, Silverpush, and Sonarax. The proposed blocking strategy has shown to be effective in most cases, except when the loudness of the sender exceeds a certain volume and thus drowns the jamming signal.

\section{Demonstration system}

\begin{figure}[ht]
\includegraphics[trim={0 0cm 0 0cm},clip,width=0.9\linewidth]{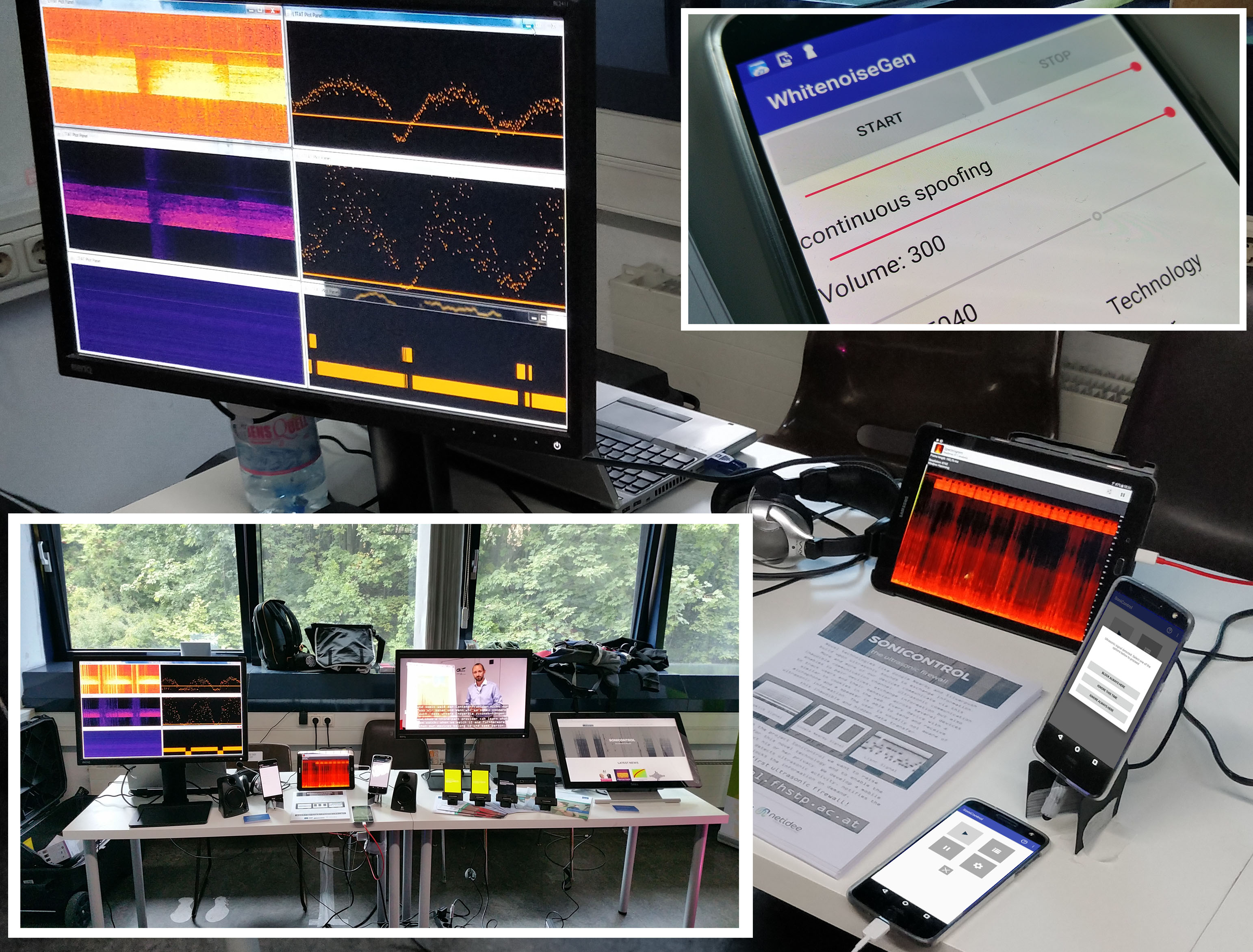} 
\caption{The real-time demonstration system. Users can participate by downloading our app from Google Play Store.}
\label{fig:demoSystem}
\end{figure}

Figure \ref{fig:demoSystem} depicts our demonstration setup, which interactively showcases first, how to send data via ultrasound between devices and second, how to detect and block ultrasonic communication effectively. Thanks to spectrogram displays, the user can see what happens during this inaudible process. The user can also participate in the demo herself by (i) sending and receiving ultrasonic information as well as by (ii) trying out the SoniControl firewall on her own device to detect and block unwanted traffic.

\section{Conclusion}
We have implemented a first ultrasonic firewall that detects different types of ultrasonic transmission technologies in real-time and provides an effective means to prevent unwanted data exchange. The app ``SoniControl'' has been recently released in Google's Play Store (until now 17000 downloads) and the source code is provided publicly to the community. As a primary goal, the app should raise awareness to novel multimedia-based data exchange practices and improve privacy protection.

\begin{acks}
  The authors thank Kevin Pirner and Peter Kopciak for their contributions. 
  This project was funded by the \emph{netidee} initiative of the Internet Foundation Austria under grant no.~1701 and 2110.

\end{acks}

\bibliographystyle{ACM-Reference-Format}
\bibliography{sample-bibliography}

\end{document}